\documentstyle[multicol,aps,prl,epsf,psfig]{revtex}
\begin{document}
\def\bea{\begin{eqnarray}}
\def\eea{\end{eqnarray}}
\def\a{\alpha}
\def\D{\langle l \rangle}
\def\p{\partial}
\draft
\title{Analytical Studies of Strategies for Utilization of {\em
Cache} Memory in Computers}
\author{Satya N. Majumdar and Jaikumar Radhakrishnan}
\address{Tata Institute of Fundamental Research, Homi Bhabha Road, 
Mumbai-400005, India}
\maketitle

\begin{abstract}
We analyze quantitatively several strategies for better utilization of
the {\em cache} or the {\em {fast access}} memory in computers. 
We define a performance factor $\alpha$  that denotes the
fraction of the cache area utilized when the main memory is accessed at
random. We calculate $\alpha$ exactly for
different competing strategies, including the hash-rehash and 
the skewed-associative strategies which were earlier analyzed
via simulations. 
\end{abstract}


\begin{multicols}{2}


The memory of a computer is organized in {\em pages.} A single page
consists of several {\em words.} As the computer performs its computations,
it reads from and writes into memory. 
When a word of main memory is accessed, the
entire page containing the word is fetched and stored in a {\em more
accessible} part of the computer's hardware called the {\em cache}, which
admits fast access. Subsequent accesses are likely to be for 
words in this page and hence the average memory access time is
considerably reduced\cite{Hwang-Briggs84}. 

Since fast memory is expensive, the cache usually holds far fewer
pages than the main memory. When a program requires a certain page
that is not already in the cache, the computer has to fetch it from
the main memory. But where in the cache should this new page be
placed?  The placements in the cache of these incoming pages must be
organized so that (i) very little time is spent in locating the page where
the word is stored in the cache and (ii) the cache area is maximally
utilized, i.e., ideally pages should not be sent back to the main
memory, when there is space left in the cache. 

To fix our notations, we assume that the
main memory has $M=2^m$ pages and the cache has $N=2^n$ pages, where
$m\gg n$. We use $m$-bit and $n$-bit 0-1 strings as addresses for pages
in the main memory and the cache respectively. The page of the main memory
corresponding to the address $a\in
\{0,1\}^m$ will be denoted by $P_a$. Similarly $Q_b$ will denote the
page in the cache corresponding to the address
$b \in\{0,1\}^n$. When an access to page $P_a$ is made in the main memory,
it has to be brought to a certain page $Q_b$ in the cache. So the question
is what is the best strategy for choosing $Q_b$ (i.e., cache
organization) such that the cache area
is maximally utilized. 

There are two extreme strategies for cache organization.
Strategy $1$, usually called 
{\em direct mapping} strategy, assigns a fixed location $Q_b$ in the cache
for each page $P_a$ of the main memory where $b$ is the first $n$ bits of
$a$. That is, each time $P_a$ is fetched into the cache, it
will be placed in page $Q_b$; if there is already a page of the main
memory residing at $Q_b$, then that page will be sent back to the main
memory (the main memory will be updated) and $P_a$ will replace it in
the cache. In this strategy when an access to a page $P_a$ is made, we
know exactly where to find it in the cache, namely $Q_b$. Thus it performs
well on point (i) as no time is wasted in searching. However it performs
poorly on point (ii) as a page can be sent out even when most of the cache
is unused.  

The second strategy, Strategy 2, usually called {\em associative mapping}
strategy, 
allows $P_a$ to reside
anywhere in the cache; if all the pages in the cache are already
occupied, one of them, chosen according to some rule (e.g.~least
recently used), will be sent back to the main memory to make space for
the $P_a$. In this strategy, when a page $P_a$ is accessed, 
determining if it
is already present in the cache can be expensive, both from the point
of view of time taken and the hardware needed to implement the
search. It has, however, its advantage in the
utilization of cache area, since no page in the cache is sent out
unless the cache is full.  

Thus Strategy 1, though preferable from the
point of view of design, is likely to be inferior to Strategy 2 in
utilization of the memory available in the cache. 
In order to improve the performance of caches, several other
strategies, which try to to combine the advantages of Strategies 1 and
2, have been proposed. In this letter, we will primarily be concerned
with three such strategies $A$, $B$ and $C$ mentioned below.
These strategies perform considerably better than Strategy 2 on point
(i), and are found to be better than Strategy 1 on point (ii).
Our goal in this Letter would be to compare quantitatively the
performances of these various strategies.

\noindent{\bf Strategy A:} This is known as {\em hash-rehash Strategy}
~\cite{Agarwal89}. In this strategy,
each page of the
main memory is allowed to reside in two locations 
$Q_{b_1}$ and $Q_{b_2}$ in the cache that are determined as follows: $b_1$
is the string of the first $n$ bits of $a$ and $b_2$ is the string of the
next $n$ bits of $a$. When a page $P_a$ from the main memory is brought
to the cache, it is first put in $Q_{b_1}$ provided $Q_{b_1}$ is empty; 
if $Q_{b_1}$ is occupied we place $P_a$
in $Q_{b_2}$. (If there is some page $P_{a'}$ residing at $Q_{b_2}$,
then it is replaced by $P_a$ and is sent back to the main memory.)

\noindent{\bf Strategy B:} This is known as 
{\em two-way skewed-associative
strategy}~\cite{Seznec-Bodin93,Bodin-Seznec95,Seznec93,Seznec97}.
In this case, we
divide the cache into
two banks $Q$ and $Q'$, each with capacity $2^{n-1}$. The pages of
the two banks are denoted by $Q_b$ and $Q'_{b}$ respectively where $b \in
\{0,1\}^{n-1}$. With each page $P_a$ are associated two pages
$Q_{b_1}$ and $Q'_{b_2}$, one from each bank. When a page $P_a$ is brought
to the cache, we first try to store $P_a$ in the
first bank at location $Q_{b_1}$, and if that fails, we store it in
the second bank at location $Q'_{b_2}$. 

\noindent{\bf Strategy C:} This is known as
{\em two-way set-associative strategy}
~\cite{Agarwal89,Hwang-Briggs84}. Here the cache is again divided into
two banks. 
The only difference is that we associate pages $Q_{b_1}$ and $Q'_{b_1}$
with page $P_a$. That is, if 
$Q_{b_1}$ is already taken, we try to store the page in the second
bank but at location $Q'_{b_1}$ (and not $Q'_{b_2}$ as in Strategy $B$).

Clearly, Strategies $A$, $B$ and $C$ are variants of Strategy $1$. 
Recently, based on data obtained from simulations, Seznec and
Bodin~\cite{Seznec-Bodin93,Bodin-Seznec95,Seznec93,Seznec97} have strongly
advocated the use of skewed-associative caches (Strategy B). However no
attempt appears to have been made to compare the performance of this
strategy with older strategies within a theoretical framework.
In this letter, we take the first steps in this direction.
In particular, we calculate analytically a quantity called
{\em performance factor} $\alpha$ (defined below) for various strategies
mentioned above.

The cache has place to store $N=2^n$ pages.  If $N$ pages
are accessed at random, a perfect strategy would accomodate them all
in the cache, without sending any page back to the main memory. But
for most practical strategies some pages will be sent back to the main
memory. We define the performance factor $\alpha$ of a strategy to be
the expected fraction of the randomly accessed $N$ pages (in the $N\to
\infty$ limit) that get accomodated in the cache. A perfect strategy
has $\alpha=1$. For all other strategies, $\alpha\le 1$. We show below
that $\alpha_1=1-e^{-1}=0.63212\ldots$ for Strategy 1 but it increases
considerably to $\alpha_A=(e^2-1)/(e^2+1)=0.76159\ldots$,
$\alpha_B=1-{1\over {2e^2}}(1+e^{1-e^{-2}})=0.77167\ldots$ and
$\alpha_C=1-2e^{-2}=0.72932\ldots$ for variants $A$, $B$ and $C$
respectively.

It turns out that $\alpha_1$ and $\alpha_C$ can be computed quite easily
using elementary probability arguments. 
Let us first compute $\alpha_1$ for Strategy 1 in its original form.
Suppose a page $P_a$ of the main memory is chosen at random. The first
$n$ bits of $a$ will be uniformly distributed in $\{0,1\}^n$. Thus,
each access corresponds to the choice of one of the $N=2^n$ pages of
the cache, and we want to know how many distinct pages of the cache
are expected to be chosen after $N$ random pages of the main memory
are accessed. In other words, we have $N$ bins and $N$ balls and each
ball is thrown at random into one of the bins; if the bin is already
occupied then the new ball replaces the existing ball. (Clearly, it
makes no difference to the calculations if we think that the old ball
stays and it is the new ball that is discarded.)  What is the expected
number of bins that will be occupied after all $N$ balls have been
tried? The probability that any fixed bin remains empty after $k$
balls have been
tried is $(1-1/N)^k$. Therefore, the expected fraction $x_1(k)$ of the
occupied bins
after $k$ trials is given by, 
\bea
x_1(k)=1-\big ( 1-{1\over {N}}\big )^k.
\label{expec1}
\eea
The performance factor is then given by,
\bea
\alpha_1=\lim_{n\to \infty}x_1(n)=1-{1\over {e}}=0.63212\ldots.
\label{perform1}
\eea

Similar arguments can be used for Strategy C also. Here there are two
banks and each bank has $N'=N/2$ bins. For every bin in the first bank,
there
is an associated bin in the second bank. A bin from the first bank is
chosen at
random and a ball is thrown into that bin. If the bin was already occupied
then the ball is thrown into the associated bin in the second bank. If the
second bin was also occupied, then the ball is rejected.
Let $x_1(k)$ and $x_2(k)$ denote the fraction of occupied
bins in the first and the second bank respectively after $k$ trials.  
Then clearly $x_1(k)=1-(1-1/N')^k$ as given by Eq. (\ref{expec1}).
We also note that the probability that a bin in the second bank
remains unoccupied given that its partner in the first bank is occupied 
is simply given by $x_1(k)-x_2(k)={k\over {N'}}(1-1/N')^{k-1}$. This is
because out of $k$ trials, only one (which can be $1$-st,
$2$-nd,$\ldots$, or the $k$-th trial) should be succesful in choosing
the given bin in the first bank and the others should be unsuccessful.
So the performance factor $\alpha_C$ is given by,
\bea
\alpha_C={\lim_{N\to \infty}}{{x_1(N)+x_2(N)}\over
{N}}=1-2e^{-2}=0.72932\ldots.
\label{performC}
\eea

It turns out that such elementary arguments, however, do not give us
the expressions for the fraction of occupied bins for Strategies A
and $B$ and one has to unfortuantely carry out more detailed computations 
for those two cases which we outline below.

We first consider Strategy $A$. In this case there are two fixed
locations $Q_{b_1}$ and $Q_{b_2}$ in the cache corresponding to a page
$P_a$ in the main memory. If $a$ is an $m$ bit string then $b_1$
consists of the first $n$ bits of $a$ and $b_2$ the next $n$ bits of
$a$ (assuming $m\gg 2n$). Suppose that an address $a$ is accessed
randomly by the computer. As $a$ varies uniformly over $\{0,1\}^m$,
the two corresponding strings $b_1$ and $b_2$ also vary uniformly over
$\{0,1\}^n$ and their distributions are independent. Thus, in the
language of balls and bins, we have the following process. There are
$N$ balls and $N$ bins. The balls are placed one after another using
the following strategy. For a given ball, a bin is chosen at random
and the ball is attempted to be placed there. If the bin is empty the
ball occupies the bin; if the bin is occupied,
another bin is picked at random (note $b_2$ is independent of
$b_1$ unlike in Strategy C). If this second bin is empty, the ball
occupies it. However, if
this bin is also occupied, then the ball is discarded. 

We define $P_{A}(r,k)$ to be the probability
that $r$ bins are occupied after ``time" $k$, i.e., after $k$ trials.
Note that in this case
each ball is tried at the most twice. The evolution equation for
$P_A(r,k)$ is then given by,
\bea
P_{A}(r,k+1)&=&\big ({r\over {N}}\big )^2P_{A}(r,k)+\big [ 1-\big     
({{r-1}\over {N}}\big )^2\big ] \nonumber \\
&\times& P_A(r-1,k).
\label{evola}
\eea
with the ``boundary" condition, $P_A(0,k)=\delta_{k,0}$ and the ``initial"
condition, $P_A(r,0)=\delta_{r,0}$. This equation can be written in a
compact matrix form, $[P_A(r,k+1)]={\hat W}[P_A(r,k)]$ where $\hat W$  
is the $(n+1)\times (n+1)$ evolution matrix whose elements can be easily
read off Eq. \ref{evola}. This equation can be solved using the
standard techniques of statistical physics which we outline below.
The general solution can be written as
\bea
P_A(r,k)=\sum_{\lambda}a_{\lambda}Q_{\lambda}(r){\lambda}^k
\label{sola}
\eea
where $[Q_{\lambda}(r)]$ is the right eigenvector (with eigenvalue 
$\lambda$) of the matrix $\hat W$ and $a_{\lambda}$'s are constants to be
determined from the initial condition. The eigenvalues of $\hat W$ are  
simply, $\lambda=l^2/N^2$, where $l=0,1,2,\ldots, N$. The $l$-th
right eigenvector satisfies the equation,
\bea
{r^2\over {N^2}}Q_l(r)+\big [ 1-{(r-1)^2\over {N^2}}\big
]Q_l(r-1)={l^2\over {N^2}}Q_l(r),
\label{eigena}
\eea
for all $r\ge 1$. The generating function ${\tilde
Q}_l(z)=\sum_0^{N}Q_l(r)z^r$
satisfies the differential equation,
\bea
z^2(1-z){ {d^2{\tilde Q}_l}\over {dz^2}}+z(1-z){ {d{\tilde Q}_l}\over
{dz}}-(l^2-N^2z){\tilde Q}_l=0.
\label{diffa}
\eea
With a little algebra, it is not difficult to show that the well
behaved solution of this differential equation is given by,
${\tilde Q}_l(z)=z^l(1-z)P_{N-l-1}^{1,2l}(2z-1)$ for $l<N$ and
${\tilde Q}_N(z)=z^N$ where $P_n^{a,b}(x)$'s are Jacobi polynomials
defined as,
\bea
P_n^{a,b}={1\over {2^n}}\sum_{m=0}^{n}{{n+a}\choose m}{{n+b}\choose
{n-m}}   
\big ( x-1\big )^{n-m} \big ( x+1 \big )^m.
\nonumber
\eea
The final solution for the generating function, ${\tilde P}_A(z,k)=
\sum_{0}^N P_A(r,k)z^r$ is given by,
\bea
{\tilde P}_A(z,k)=z^N+\sum_{l=0}^{N-1}a_l
z^l(1-z)P_{N-l-1}^{1,2l}(2z-1)\big
( {l\over {N}}\big )^{2k}
\nonumber
\eea
where the coefficients $a_l$'s are determined from the initial condition
${\tilde P}_A(z,0)=1$. After some amount of algebra, we get,
\bea
a_l=(-1)^{l+N-1}{ {n(2-\delta_{m,0})}\over {n-m}}
\eea
for $l=0,1,2,\ldots N-1$. Thus the expected fraction of occupied bins
after time
$k$, $x_A(k)={1\over {N}}{{\partial {\tilde
P}_A(z,k)}\over {\partial z}}|_{z=1}$ is given by,
\bea
x_A(k)= 1+2\sum_{l=1}^{N-1}(-1)^{l+N}\big ( {l\over {N}}\big )^{2k}.
\label{finala}
\eea
Putting $k=N$ in Eq. \ref{finala} and taking $N\to \infty$ limit, we
get the performance factor,
\bea
\alpha_A= {{e^2-1}\over {e^2+1}}= 0.76159\ldots .
\label{performa}
\eea

Next, we consider Strategy B. In this case, the cache area is divided
equally into two banks. Corresponding to every page $P_a$
in the main memory, there are two fixed locations $Q_{b_1}$ in
the first compartment and $Q_{b_2}$ in the second. 
Again $b_1$ and $b_2$ are uniformly and
independently distributed over $\{0,1\}^{n-1}$. The operation on the cache
then corresponds to the following problem with balls and bins. There
are $N=2^n$ bins, half of which, $N'=2^{n-1}$, belong to the
first compartment and the remaining $N'$ belong to the second. There
are $N$ balls
and each is placed in the bins according to the following rules. One
of the first $N'$ bins is chosen at random. If the bin is occupied
one of the $N'$ bins in the second half is chosen. If even this bin is
occupied the ball is discarded. 

Let $P_B(r_1,r_2,k)$ denote the probability
that after ``time' $k$, the first
compartment has $r_1$ occupied bins and the second $r_2$ occupied bins.
Now the second compartment is tried only if the first compartment fails
to accomodate a given ball. If there are $r_1$ occupied balls in the first
compartment at time $k$, clearly the scond compartment has been tried
only $(k-r_1)$ times, out of which $r_2$ attempts have been successful.
Noting that for each compartment separately, the local strategy is exactly
as in Strategy 1 defined earlier, it is easy to see that,
\bea
P_B(r_1,r_2,k)= P_1(r_1,k)P_1(r_2,k-r_1)
\label{prodb}
\eea
where $P_1(r,k)$ is the probability of having $r$ occupied balls in the
first compartment after $k$ trials and is the same as in Strategy 1.
$P_1(r,k)$ can be computed exactly by following the same steps as used
for Strategy $A$. Using this exact result in Eq. \ref{prodb} and after
some algebra, we finally
find the expected fraction of occupied bins (taking into account both
the banks) after $k$ trials,
\bea
x_B(k)=1-{1\over {2}}\big (1-{1\over {N'}}\big
)^{k} S(N',k)
\label{lbutb}
\eea
where $S(N',k)$ is the sum,
\bea
S(N',k)=\sum_{l=0}^{N'}(-1)^{N'-l}{N'\choose l}
{{N'}^{l}\over
{(N'-1)^{N'}}}\big ( { m\over {N'}}\big )^{k}.
\nonumber
\eea
Putting $k=N$ and taking the limit $N\to \infty$, we finally get
the performance factor $\alpha_B$ for Strategy B,
\bea
\alpha_B= 1- {1\over {2e^2}}\big (1+e^{1-e^{-2}}\big )=0.77167\ldots .
\label{performb}
\eea

The method illustrated above can not only calculate
the final performance factor but it also gives exactly the full
probability 
distribution of the number of occupied bins at any arbitrary ``time"
$k$ and for any cache size $N$. It turns out that there is an easier
method (to be described below) to compute just the performance factor
and is valid only in the $N\to \infty$ limit.
It not only 
correctly reproduces the exact results for $\alpha$'s as found above but
can further be used to compute the performance factors for more
general and complicated strategies.

Consider, for example, a generalized version of Strategy $A$ where instead
of $2$
fixed locations per page in the cache, one now has $p$ fixed locations.
In the ``ball and bin" language, each ball is tried at most $p$ times
before being discarded.   
Let $r(k)$ be the random variable that denotes the number of occupied bins
after discrete ``time" $k$. Clearly, the increment $\Delta
r(k)=r(k+1)-r(k)$ is
also a random variable that takes the value $0$ with probability
$(r/N)^p$ (when all $p$ trials for a given ball are unsuccessful) and $1$
with probability $1-(r/N)^p$. Taking expectations, we get
$\langle r(k+1)\rangle - \langle r(k)\rangle =1-\langle (r/N)^p\rangle$.
We then define $x=r/N$ and $t=k/N$ and note that they become continuous
in the $N\to \infty$ limit and $\langle x\rangle$ evolves as
\bea
{ {d\langle x\rangle}\over {dt}} =1-{\langle x^p\rangle }.
\label{avg1}
\eea
Using the {\em method of bounded differences}~\cite{McDiarmid}, it can
be shown that for each $t\in [0,1]$ and $\epsilon >0$
\bea
\Pr[|x(t) - \langle x(t)\rangle| \geq \epsilon] ~\leq 2
\exp(-2\epsilon^2N/p).
\nonumber
\eea
From this, it follows that
$|{\langle x^p\rangle}-{\langle x\rangle}^p| \leq O(N^{-{1 \over
2}})$. Thus, one can 
neglect fluctuations in the $N\to \infty$ limit. Using this in
Eq. \ref{avg1}, one gets a closed equation for $\langle x\rangle$($t$)
which can then be integrated to give the performance
factor, $\alpha_p= \langle x\rangle$($t=1$). For general $p$, we get
$\alpha_p$ as solution of the equation,
\bea
\int_0^{\alpha_p}{dy\over {1-y^p}}=1.
\label{galpha}
\eea
For example, for $p=1$ and $p=2$, we reproduce respectively,
$\alpha_1=1-e^{-1}$ and $\alpha_A=(e^2-1)/(e^2+1)$ from Eq.
\ref{galpha}. For large $p$, we find from Eq. \ref{galpha},
\bea
\alpha_p=1- {{\log 2}\over {p}} + O\big ( {1\over {p^2}}\big ).
\label{largep}
\eea
Thus as $p$ increases, the deviation from perfect behaviour $(1-\alpha_p)$
decays as a power law $\sim 1/p$.

In a similar fashion, one can consider a generalized
version of Strategy $B$ where instead of 2 equal compartments in the
cache, one now considers $p$ equal compartments. In this case, once again
the performance factor for general $p$ can be easily computed using the
continuous time method. Apart from reproducing the result for $p=2$,
we find that for large $p$, 
\bea
\alpha_B(p)= 1-{1\over {(e-1)p}} +O \big ( {1\over {p^2}}\big ).
\eea
Another generalization of Strategy $B$ would be to consider $2$
banks only but of unequal sizes $N$ and $\gamma N$ respectively.
Then a similar calculation shows that the performance factor, which
is now a function of $\gamma$, is given by,
\bea
\alpha_B(\gamma)=1-{{e^{-(1+\gamma)}}\over {1+\gamma}}-{{\gamma}\over
{e(1+\gamma)}}e^{-{\exp {[-(1+\gamma)/{\gamma}]} }}.
\label{Gamma}
\eea
This function has a single maximum at $\gamma=0.68932\ldots$ where
$\alpha_B(max)=0.775862\ldots$, clearly better than
$\alpha_B(1)=0.77167\ldots$, the result for two equal banks.

In summary, we have studied analytically the performance of several
strategies for cache utilization. We have derived exact values for
cache utilization factors for these strategies under the assumption that
pages in the main memory are accessed at random.  

Cache utilization is just one of several issues in the design of
caches. Indeed, it has been reported~\cite{Agarwal89,Seznec93} that
two-way set associative caches require smaller execution times
than hash-rehash caches,
although, our analysis indicates that the latter utilize the cache
better. 

To get performance values that will be useful to designers of cache,
one should take into account the amount of time spent in decoding
addresses, updating tables and the cost of hardware. Also,
instead of assuming that the pages of the main
memory are accessed at random, a better understanding of the access
patterns in typical applications might be needed.

We are grateful to Abhiram Ranade for bringing the question of analysis of
skewed associative caches to our attention; in particular, the idea of
using unequal banks in this strategy is due to him. We also thank
D. Dhar for useful discussions.

\end{multicols}

\end{document}